\documentclass[aps,prl,reprint,superscriptaddress]{revtex4-2}   

\usepackage{graphicx}
\usepackage{dcolumn}
\usepackage{bm}
\usepackage{amsmath}
\usepackage{natbib}
\usepackage{verbatim}
\begin{document}
\title{\textbf Modulating Effect of Evanescent Waves on Thin-Film Growth}
\author{Rongjing Guo}
\affiliation{Institute of High Energy Physics, Chinese Academy of Sciences, Beijing 100049, China}
\author{Tai-Chang Chiang}
\affiliation{Department of Physics, University of Illinois at Urbana-Champaign, Urbana, Il 61801, U.S.A}
\affiliation{Frederick Seitz Materials Research Laboratory, University of Illinois at Urbana-Champaign, Urbana, Il 61801, U.S.A}
\author{Huan-hua Wang}
\email[Correspondence: ]{wanghh@ihep.ac.cn}
\affiliation{Institute of High Energy Physics, Chinese Academy of Sciences, Beijing 100049, China}
\affiliation{School of Nuclear Science and Technology, University of Chinese Academy of Sciences, Beijing 100039, China}
\date{Received Dec. 28, 2020; Revised July 26, 2021}

\begin{abstract}
Atomic-scale smooth thin films are keys to successful integration and proper function of many multilayer-structured devices. However, the intrinsic island-like growth mode prevents human being from realizing such ultrasmooth films of many important functional materials. To solve this problem, we propose a negative entropy-infusing method that employs evanescent waves to enhance the downward interlayer diffusion of adatoms and thus transform the island-like growth mode to the layer-by-layer growth mode. The formulas of the optical force and the lowered diffusion barrier were derived, and the application of this theory on a simplified example demonstrates significantly improved surface morphologies through numerical simulations.
\end{abstract}

\keywords{Thin films; Growth modes; Surface morphology; Evanescent wave; Downhill diffusion}

\maketitle
\section{\romannumeral1. Introduction}
For most integrated optoelectronic devices which function on the basis of heterojunctions, the sharpness of the interfaces is a key parameter determining the quality of the devices \cite{1}. The fabrication success rate, consistency and working life of heterojunction devices are all related to the surface roughness of the component thin films, so controlling the surface roughness of component thin film is one of the key factors for optimizing heterojunction devices. But as a complex phenomenon, surface morphology of growing thin films is determined by many physics factors (non-uniform distribution in flux density, energy and momentum of incident species, kinetic energy, momentum, growth temperature, growth rate, sticky coefficient, diffusion, Ehrlich-Schwoebel barrier (ESB) \cite{2,3,4} and so on) and physics effects (the random distribution effect, the anistropic growth rate effect, surface energy effect, strain effect, shadowing effect \cite{5}, quantum size effect \cite{6,7}, steering effect \cite{8}, etc.). Optimizing all of the showing-up factors is a quite time-consuming solution usually adopted for achieving ultra-smooth surface, but this solution may not work for many complex compound functional materials.

In fact, the applications of surface morphologies of thin films can be classified into two categories. One category is utilizing rough surfaces, which can be achieved using many methods such as surface etching, glancing-angle deposition \cite{5}, and various other surface nanostructure fabrication techniques. The other category utilizes atomic-scale smooth surfaces or interfaces, which cannot be always satisfied because not every material can grow in a layer-by-layer mode. For example, to fabricate sandwich-type Josephson junctions using two YBa$_2$Cu$_3$O$_{7-\delta}$ (YBCO) layers sandwiching an insulating layer between, the middle layer must be both insulating and of a uniform thickness around the superconducting coherence length (13.5${\pm}$1.5\AA{} along a,b-axis) of YBCO simultaneously \cite{9}, which means at least one YBCO layer and the insulating layer must have atomic-scale smooth surfaces. However, YBCO thin film does not grow in a layer-by-layer mode, so high-temperature superconducting computers based on YBCO have been unavailable so far. Therefore, if we invent a universally usable modulating method of altering thin-film growth mode from the island-like (Volmer-Weber) growth to a 2-dimensional (layer-by-layer or step-flow) mode, we can not only open up new research fields that rely on atomic-scale smooth thin films, but also realize the industrial applications of important functional materials and thus greatly promote new technology developments, which is of epoch-making significance.

\begin{figure}
	\includegraphics[width=8cm]{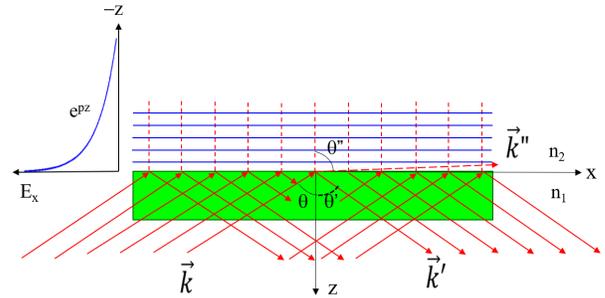}
	\caption{Schematic of EW in optically thinner medium. The horizontal blue lines represent the isoamplitude planes, and the vertical red lines the isophase planes. Light refraction at the vacuum-substrate interfaces is ignored for simplicity.}
\end{figure}

To achieve atomic-scale smooth surface of a thin film material growing intrinsically in an island-like way, its growth mode must be altered into 2D growth manner. A basic train of thought is that considering the surface roughness is a kind of disorder (surface entropy) caused in a large degree by the random surface diffusion, we need infuse negative entropy into surface to break the randomness of adatom diffusions on the growing nuclei or islands for decreasinge surface roughness. In this article, we come up with a method of using evanescent waves (EW) to enhance the downward interlayer diffusion and suppress the upward interlayer diffusion, thus smoothen the surface of a growing thin film. The article is organized as follows: First, we introduce the main idea of the modulating effect of EW on a growing thin film. Then, we derived the formulas of the optical force exerted on adatoms by EW and the lowered part of the diffusion barrier. In the following, to illustrate the feasibility of this method using an example, the diffusion barriers without EW and with EW of different amplitudes were simulated on Ag thin film, and to visualize the modulating effect on its surface morphology, Kinetic Monte Carlo simulations were performed based on the quasi-minimum energy paths (quasi-MEPs) and ESBs of Ag(100) surface obtained from DFT calculations. Finally, the ubiquitousness of this method was discussed. The significant enhancement of downward interlayer diffusion by EW theoretically verify the feasibility of this method, and experimental tests are suggested using time-resolved x-ray scattering (trXRS) \cite{10} and grazing-incidence x-ray photon correlation spectroscopy (GIXPCS) \cite{11}.

\section{\romannumeral2. Principles}

It is well known that total reflection produces evanescent wave field in the optically thinner medium \cite{12}. As depicted in Fig.1, when a laser beam is incident onto a thin film from the substrate side with an incident angle $\theta$ larger than the critical angle $\theta_{c}$, total reflection happens at the thin film-vacuum interface and an EW field is generated at the vacuum side near the thin film surface. The electric field of EW parallel to the reflection plane is a strong gradient field which can polarize atoms and exert a net attraction force pulling the polarized atoms downwards to the interface. Therefore we can use this attraction force to enhance the downward diffusion and suppress the upward diffusion of the adatoms on the nuclei or islands of a growing thin film, which should result in a smoothened surface.

According to fundamental optics principles, the electric field of EW can be derived
\begin{equation}
E_{EW} = E_0e^{-kz\sqrt{sin^2\theta - \epsilon_1/\epsilon_2}}e^{i(\omega t - kxsin\theta )}
\end{equation}
where $\omega$ is angular frequency, $k$ is the wave vector of incident wave, $\theta$ denotes the incident angle, and $\epsilon_1$,$\epsilon_2 $ represent the dielectric constants of Medium 1 and 2, respectively. The direction of the electric field depends on its polarization state.

Based on Lorentz effective field theory \cite{13,14,15}, the local electric field at position $\boldsymbol{r_i}$ occupied by an atom can be obtained by summation of the external electric field at this site and the fields induced by dipole oscillators around the atom, which can be expressed as \cite{13,16}
\begin{equation}
\boldsymbol{E_{loc}}(\boldsymbol{r_i},\omega) =  \boldsymbol{E_{ex}}(\boldsymbol{r_i},\omega) + \sum_{i \neq j} \boldsymbol{E_{dip}}(\boldsymbol{r_i},\boldsymbol{r_j},\omega)
\end{equation}
\begin{equation}
\boldsymbol{E_{dip}}(\boldsymbol{r_i},\boldsymbol{r_j},\omega) = \frac{3(\boldsymbol{p}(\boldsymbol{r_j},\omega) \cdot \boldsymbol{r_{ij}}   )\boldsymbol{r_{ij}}-r_{ij}^2 \boldsymbol{p}(\boldsymbol{r_j},\omega) }{4 \pi \epsilon_0 r_{ij}^5}
\end{equation}
where $\boldsymbol{E_{ex}}(\boldsymbol{r_i},\omega) $ is the external electric field at position $\boldsymbol{r_i}$, $\boldsymbol{p}(\boldsymbol{r_j},\omega)$ denotes the dipole moment at position $\boldsymbol{r_{j}}$, and $\boldsymbol{E_{dip}}(\boldsymbol{r_i},\boldsymbol{r_j},\omega)$ represents the electric field at $\boldsymbol{r_i}$ induced by the dipole oscillator at $\boldsymbol{r_j}$. Here, the retardation effect is ignored due to the small separation $r_{ij} $ between two adatoms, i.e. quasi-static approximation.

Let $\alpha(\boldsymbol{r_i},\omega)$ be the polarizability of adatom at $\boldsymbol{r_i}$,  the dipole moment $\boldsymbol{p}(\boldsymbol{r_i},\omega)$ at that position can be obtained
\begin{equation}
\boldsymbol{p}(\boldsymbol{r_i},\omega) = \alpha(\boldsymbol{r_i},\omega) \boldsymbol{E_{loc}}(\boldsymbol{r_i},\omega)
\end{equation}
\begin{equation}
\alpha(\boldsymbol{r_i},\omega) = \alpha_0(\boldsymbol{r_i},\omega)/[1-(2/3)ik_0^3\alpha_0(\boldsymbol{r_i},\omega)]
\end{equation}
where $k_0 = \omega/c$ and $\alpha_0(\boldsymbol{r_i},\omega)$ is given by the Clausius-Mossotti relation:
\begin{equation}
\alpha_0(\boldsymbol{r_i},\omega) = 3 a^3 \epsilon_0 \frac{\epsilon(\omega)-1}{\epsilon(\omega)+2}
\end{equation}
where $a$ represents the size of adatoms.

To obtain the self-consistent solutions of $\boldsymbol{p}(\boldsymbol{r_i},\omega)$ of dielectric grains with arbitrary shape, an iterative method can be employed. Then, using Maxwell tensor method we can obtain the optical force on atom at position $\boldsymbol{r_i}$ \cite{12}
\begin{equation}
\boldsymbol{F_{opti}}(\boldsymbol{r_i},\omega) = [\boldsymbol{p}(\boldsymbol{r_i},\omega)\cdot \nabla]\boldsymbol{E_{ex}}(\boldsymbol{r_i},\omega)
\end{equation}

\begin{figure}
\includegraphics[width=6cm]{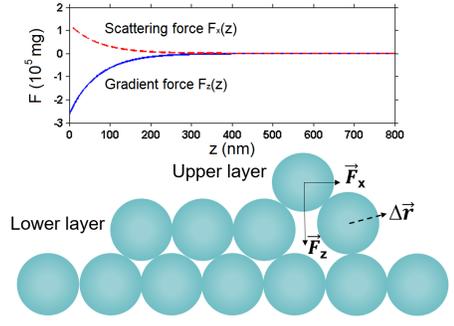}
\caption{Bottom: Schematic of interlayer diffusion in exchange process enhanced by the gradient force $\vec{F}_z$ and scattering force $\vec{F}_x$ arising from EW. $\Delta\boldsymbol{\vec{r}}$ indicates the moving direction of the squeezed atom in the lower layer. Upper: The z dependence of $F_z$ and $F_x$ for Ag adatoms on $SrTiO_3$ substrate with a 50 watt 440 nm blue laser focused down to a 10 micron spot size.}
\end{figure}
As illustrated in the bottom part of Fig.2 where Ag adatoms on $SrTiO_3$ is used as an example, the optical force include two components, one is gradient force (GF). Its direction  points to the interface, pulling adatoms to move downwards, favorable to downward interlayer diffusion. Because the light experiences scattering by the adatoms and energy loss during propagation, EW also generates a scattering force (SF) along its propagating direction.\cite{17}. The upper part of Fig.2 shows the z-dependence of GF and SF, calculated using real data of Ag adatoms on $SrTiO_3$ substrate [18,19] with a 50 watt 440 nm blue laser focused down to a 10 micron spot size. Calculations shows that the optical forces are much larger than the gravities and similar z-dependence curves also exists for other kinds of metal adatoms, as shown in Fig.1S in the Supplemental Material.

The surface morphology of a growing thin film is determined by many atomic processes mainly including deposition, intralayer diffusion, interlayer diffusion, nucleation and re-evaporation. Among them interlayer diffusion plays the most important roles in shaping surface morphology, which processes in two ways. One is that the diffusing atoms hop over the step to the neighboring layers, which is called hopping mechanism; the other is that the atoms on the step squeeze into the lower layer, and the nearest neighbor atoms in the lower layer diffuse, which is called exchange mechanism. Because a hopping adatom must experience a configuration of lower coordination number (CN) that corresponds to a higher diffusion barrier, the exchange process usually is more favorable than the hopping process, which is supported by many researches.\cite{20,21,22} Therefore, at least for the growth of metal thin films the exchange mechanism is the dominant process in interlayer diffusions. Consequently, this process can represent real interlayer diffusions for verifying the effectiveness of the modulating effect of EW on thin-film growth.

We first discuss the adatom diffusion under no EW. In a diffusion process, an adatom diffuses to the nearest-neighboring position with diffusion rate $v$. According to Boltzmann statistics and Arrhenius formula, the diffusion rate $v$ can be expressed as
\begin{equation}
v= v_0 \, exp(- \frac{E_{d}}{k_{B}T})
\end{equation}
where $k_{B}$ is Boltzmann constant, $v_0$ is the vibrational frequency, $T$ is the temperature, and $E_{d}$ is the diffusion energy barrier.

With EW field applied, the gradient force of EW pulls the adatoms on the step edges to squeeze into the lower layer more readily while in upward diffusion the adatoms need to overcome the GF by doing extra work, therefore the downward diffusions get promoted and the upward diffusions are suppressed. Because the light frequency is much higher than the vibrational frequency of adatoms that is about $10^{12}$ to $10^{13} s^{-1}$\cite{1,23}, the optical force during a diffusion move can be averaged into an invariable local optical force field (LOFF) $\epsilon_{loc}(\boldsymbol{r}) $. With setting the position of zero potential energy at the initial site of the diffusing adatoms, we can calculate the diffusion barriers via $ab$-$initio$ simulations, and in the case with EW, the LOFF potential energy should be appended to $E_{d}$.
\begin{equation}
E^{\prime}(\boldsymbol{r}) = E_d(\boldsymbol{r}) + \epsilon_{loc}(\boldsymbol{r})
\end{equation}
where $\boldsymbol{r}$ represents the atomic position, and $\epsilon_{loc}(\boldsymbol{r}) $ can be obtained from our theoretical model with atomic diffusion path
\begin{equation}
\epsilon_{loc}(\boldsymbol{r}) = -\int_{Path(\boldsymbol{r})}^{} \left\langle \boldsymbol{F^{\mu}_{opti}} \right\rangle \boldsymbol{dr}
\end{equation}
Therefore, the diffusion rate can be rewritten as
\begin{equation}
v^{\prime}= v_0 \, exp[- \frac{E^{\prime}(\boldsymbol{r})_{peak}}{k_{B}T}]
\end{equation}
where $E^{\prime}(\boldsymbol{r})_{peak}$ is the peak of the energy curve (the energy of initial state was set as zero), that is, the new diffusion barrier with EW field applied. Eq.(9) and Eq.(10) are valid in calculating both the ESBs and the normal intralayer-diffusion barriers.
\begin{figure}
	\includegraphics[width=9cm]{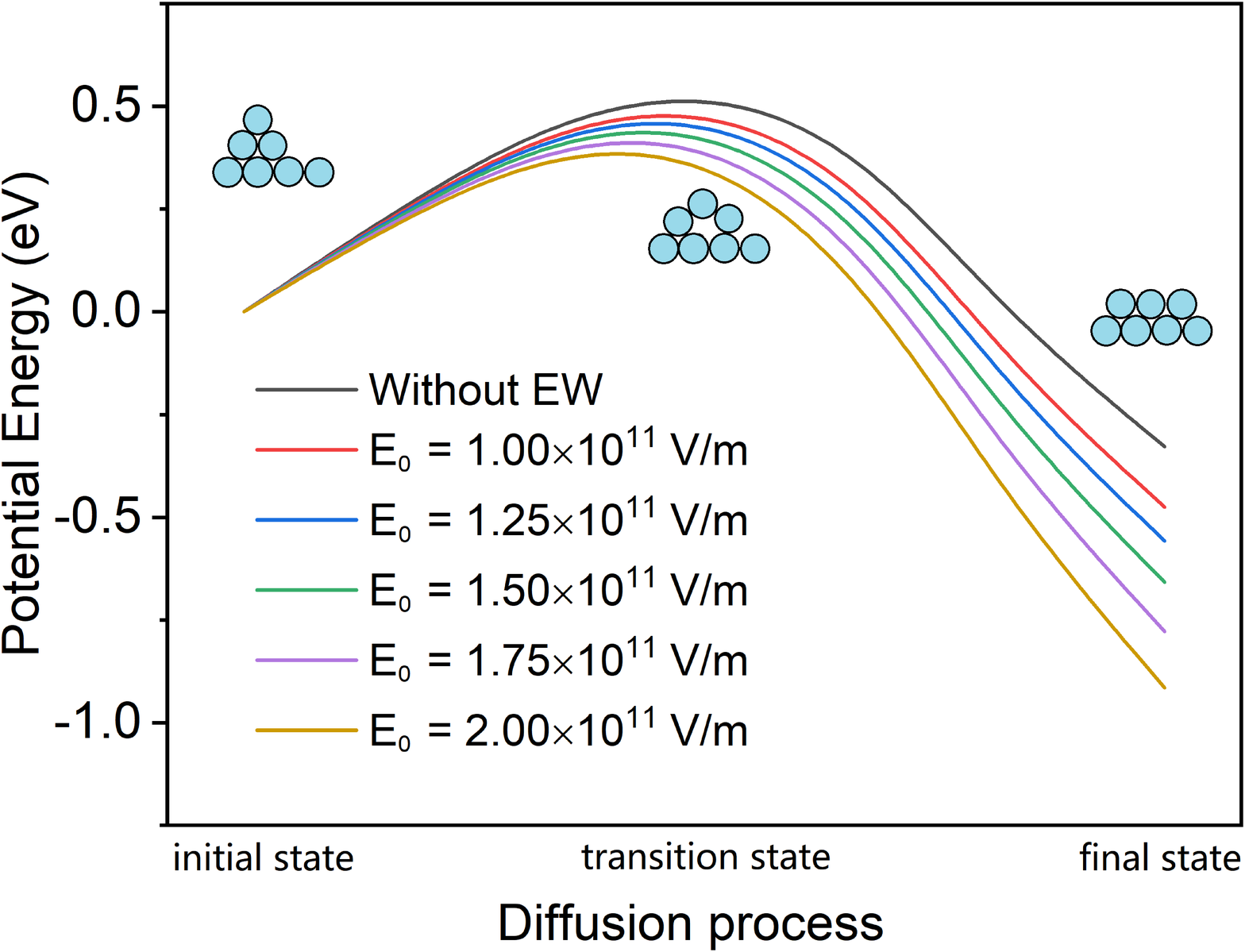}
	\caption{Interlayer diffusion barriers of Ag adatoms on (100) surface without and with EWs arising from incident light with different wave amplitudes $E_0$. To make the difference more distinct, large amplitudes of EW ranged from $1.00\times 10^{11} $ V/m to $2.00\times 10^{11}$ V/m are used in calculations. }
\end{figure}

\section{\romannumeral3. Numerical experiments}

To demonstrate the modulating effect of EW on adatoms diffusions using an example, the diffusion barriers without EW and with EW of different amplitudes were simulated on the growth of Ag thin film. To further simplify the simulations meanwhile without losing generality, we use the dominant exchange process as the representative of real interlayer diffusions to examine the enhancing effect of EW on interlayer diffusions. Therefore no other processes such as hopping across steps, edge diffusion or detachment from island edges were included in the simulations (according to the physics picture of EW-adatom interaction, the field of EW also enhance the net downward crossing-edge diffusion as explained in the following section). Our numerical experiments combine the above theoretical formulas and the DFT simulations that are performed using plane wave-based Vienna $ab$ $initio$ simulation package (VASP) \cite{24} for the purpose of finding MEPs of diffusions. We adopted the Perdew-Burke-Ernzerhof form of the generalized gradient approximation for exchange-correlation functional \cite{25}. Kohn-Sham wave functions are described by a plane wave basis est with cut-off energy of 400 eV. The thickness of vacuum layer is set up to 17 angstrom to eliminate the influence from the periodic pattern in the direction perpendicular to the surface. The diffusion paths and barriers for adatoms are evaluated using Nudge Elastic Band (NEB) method \cite{26,27}.

We selected a 3$\times$3 supercell of Ag(100) films with a K-mesh of 3$\times$3$\times$1. The initial sites of adatoms were set on the upper monolayer (ML), and the process of their interlayer diffusions from the upper ML to the adjacent lower ML were simulated. According to the diffusion path from the DFT calculations, the potential energy landscape modified by the EW effect was obtained from $Eq.(9) $ and $Eq.(10) $, as depicted in Fig.3. It is obvious that EW field lowers the downward interlayer diffusion barrier of adatoms, which is favorable to downward interlayer diffusion in the dominant exchange process, and the larger the incident light wave amplitude $E_0$ in Eq(1) becomes, the lower the barrier is.

The calculated diffusion barrier without EW is 0.54 eV, which is consistent with the previous studies (0.52 eV) \cite{21}. With EW applied, the ESB decreases significantly, for example, an incident beam of $E_0 =2.00 \times 10^{11}$ V/m can lower the ESB of the interlayer diffusion along the direction of SF by $26\%$, as Fig.3 shows. More details of the variations of barriers can be found in Table.1S of Supplemental Material. More significantly, the energy of final state is lowered more compared with no EW situation, which means the possibility of diffusing from lower layers to upper layers tend to zero, which further enhances the 2-dimensional growth behaviors. Although the GF also shrinks the interlayer spacing and alters the MEPs by reshaping the surface potential landscape, fortunately the GF is too small to change the MEPs noticeably, which makes it reasonable to take the calculated MEPs under no EW as the quasi-MEPs under EW in the calculations based on Eq.(9) and Eq.(10).
\begin{figure}
	\includegraphics[width=8cm]{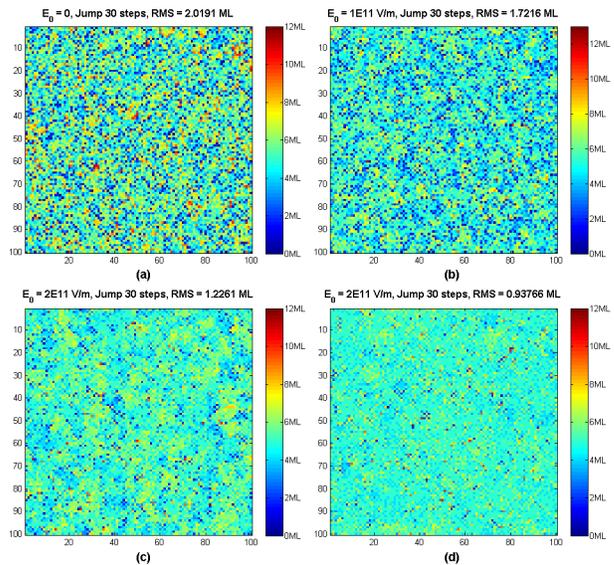}
	\caption{Surface morphology of Ag(100) thin film simulated using Kinetic Monte Carlo with 30 jumping diffusion steps. (a) without EW; (b) with both GF and SF of EW at $E_0$=1.00$\times$$10^{11}$ V/m; (c) with both GF and SF of EW and (d) with mere GF of EW at $E_0$ =2.00$\times$$10^{11}$ V/m. The simulated thin-film area is 100$\times$100 lattice points.The RMS roughnesses are listed in the titles, respectively.}
\end{figure}

To visualize the modulating effect of EW, we carried out Kinetic Monte Carlo simulations of Ag(100) thin-film growth based on the results of quasi-MEPs. Fig.4 illustrates the simulation results using the simplified model to grasp the main physics, in which 50000 deposited adatoms land on an area of 100$\times$100 lattice points, then are forced to diffuse with 30 jumping steps; when adatoms diffusing across the boundaries, periodic diffusion condition is used, i.e., the adatoms diffusing out of the simulated deposition area from one side enter the area from the other side. The striking comparison between the roughnesses of films with and without EW demonstrates its effectiveness. The details of the simulations can be found in Supplemental Material. It is noteworthy that the unidirectional nature of SF lead to the asymmetry of diffusion and thus affects smoothness of thin film. Compared with the effect of both the GF and the SF of EW, mere GF works better for smoothing surfaces, as shown in Fig 4.(d). The SF can be cancelled using two symmetrically incident beams in practical experiments.

\section{\romannumeral4. Discussion}

Because EW field can polarize any adatom and thus exert gradient force to enhance the downward interlayer diffusion, from this principle it can be concluded that this approach works universally for any material. This conclusion is very logic and solid although it has not been verified by comprehensive simulations or real experiments on the growth of complex compound thin films. It can be imagined that at any delicate near-balance state between downward and upward interlayer diffusions, even a small force can break their original quasi-balance. This should hold not only for exchange process but also for hopping process across steps, because, although the GF increases the ESB a little for adatoms hopping downward, it lowers the energy of the system with adatoms at the lower layers more since the upward displacement of an adatom in a downward crossing-step hopping process is much smaller than its downward displacement (as shown in Fig.2), and thus upward diffusing adatoms at lower layers have to overcome a larger barrier with EW applied. As a result, the EW-caused reduction in downward hopping rate can be over-compensated by the increased suppression of upward diffusion by EW. Therefore, the net downward interlayer diffusion is increased by EW in hopping process as well. Besides, in our model simulations, only 30 jumping diffusion steps have already produced obvious improvements of surface morphologies. In real cases, the trial jumping frequency around 1$\times$$10^{13}$ Hz can generate much more jumping steps even within 0.1 second, which would make the actual smoothening effect of EW much better than that in our model calculations.

\section{\romannumeral5. Conclusion}

In conclusion, the theory of the modulating effect of evanescent waves on thin-film growth is created with the formulas of the optical force and the lowered diffusion barrier of adatoms in the EW field being derived. Application of the theory on a thin film-on-substrate model using silver growth as an example demonstrates the increased net downward interlayer diffusion rate of adatoms. The simulations illustrate that EW can smoothen the surface of a growing thin film. To the best of our knowledge, this modulation method for enhancing surface flatness has not been tried hitherto. Experiments combining trXRS with GIXPCS are suggested to test this modulating effect.

\begin{acknowledgments}
This research is supported by Chinese Academy of Sciences under agreement CAS No. 11U153210485 and by NSFC of China under Contract No. Y611U21. TCC is supported by the U.S. Department of Energy (DOE), Office of Science, Office of Basic Energy Sciences, Division of Materials Science and Engineering, under Grant No. DE-FG02-07ER46383.
\end{acknowledgments}


\end{document}



\title{Modulating Effect of Evanescent Wave on Thin-Film Growth: Supplemental Material}


\author{Rongjing Guo}
\affiliation{Institute of High Energy Physics, Chinese Academy of Sciences, Beijing 100049, China}

\author{Tai-Chang Chiang}
\affiliation{Department of physics, University of Illinois at Urbana-Champaign, Urbana, Il 61801, U.S.A}
\affiliation{Frederick Seitz Materials Research Laboratory, University of Illinois at Urbana-Champaign, Urbarna, Il 61801, U.S.A}

\author{Huan-hua Wang}
\email[Correspondence: ]{wanghh@ihep.ac.cn}
\affiliation{Institute of High Energy Physics, Chinese Academy of Sciences, Beijing 100049, China}
\affiliation{School of Nuclear Science and Technology, University of Chinese Academy of Sciences, Beijing 100039, China}
\date{\today}


\date{\today}

\begin{abstract}
\end{abstract}


\maketitle

\section{Kinetic Monte Carlo}

To visualize and compare the surface morphologies of growing thin films under different growth conditions without and with the effect of evanescent waves (EW), we performed Kinetic Monte Carlo model simulations. Considering the growth process of complex materials (such as $YBa_2Cu_3O_{7-\delta}$) is too complicated and their detailed knowledge such as growth units and binding energies of various kinks and steps and edge dents is still not fully revealed, we chose the simple silver thin film as an example to simplify the simulations while grasp the main physics. In fact, even for Ag thin films, the completeness of the simulations in real details has to be sacrificed, although we use as real as possible parameters of silver for simulations.

Without losing convincing, the solid-on-solid model \cite{1}, in which no overhangs are allowed on the simulated surface, was used in the Kinetic Monte Carlo simulations. Considering the purpose of these simulations is to visualize the modulating effect of EW on thin-film growth rather than to pursue the completeness of the simulations, the simplified model simulations include only three procedures:(1) absorption of gas-phase atoms. (2) diffusion of adatoms. (3) nucleation of adatoms and formation of thin film. And the pseudo code of this method is listed in Algorithm 1. In this pseudo code, surface morphology of thin film is denoted by a 2D array $\boldsymbol{S}$. The parameters used in the Kinetic Monte Carlo simulations are shown in Table 1. These parameters of barriers were calculated based on the DFT results and Eq.(10) in the main text.

The degree of randomness of interlayer diffusions has been broken by the EW field in our simulation code. But to weaken the randomness of adatom’s intralayer diffusions, the difference of intralayer diffusions of adatoms around the step edges of nucleus or islands with in-plane bonding steps (CN=1 and 2), step kinks (CN=3), step edge dents (CN=4) and so on must be taken into accounts in the simulation code as well, which makes the simulation code too complicated while helps little to the clarification of the essential physics. So we ignored the variation of in-plane bonding of adatoms with different neighbors in our simulations, which results in the fractal-like morphologies around the islands as shown in Fig.4.In the simulated growth, the code forces bonded adatoms to diffuse in the same way that isolated adatoms diffuse, which consequently lead to a very small percentage of adatoms leave their more stable positions (bonded with neighbors) on the surface, which is the small cost paid by using simple code.

\begin{table*}[h]
	\centering
	
	\begin{ruledtabular}
		\begin{tabular}{lcdr}
			\textrm{\textbf{Algorithm 1:} Kinetic Monte Carlo}\\
			
			\colrule
			\textbf{Input:}\\
			\qquad\qquad Intralayer diffusion barriers at different directions: \\
			\qquad\qquad\qquad$bn1$(follow the direction of scattering force)\\
			\qquad\qquad\qquad$bn2$(perpendicular to the direction of scattering force) \\
			\qquad\qquad\qquad$bn3$(against the direction of scattering force) \\
			\qquad\qquad ES barriers at different directions:\\
			\qquad\qquad\qquad $bes1$(follow the direction of scattering force)\\
			\qquad\qquad\qquad$bes2$(perpendicular to the direction of scattering force) \\
			\qquad\qquad\qquad $bes3$(against the direction of scattering force) \\
			\qquad\qquad Number of particles: $N$ \\
			\qquad\qquad Step length of Diffussion: $l$\\
			\qquad\qquad Size of the lattice: $L$\\
			\qquad\qquad Temperature of substrate: $T$.\\
			\textbf{Output:} \\
			\qquad \qquad Surface morphology of thin film: $\boldsymbol{S}$. (Type of $\boldsymbol{S}$ is $L\times L$ 2d array)\\
			\\
		
			\qquad \textbf{for} $i$ in $range(0,N)$ \textbf{do:}\\
			\qquad \qquad Randomly select one site $(x,y)$ on the lattice.\\
			\qquad \qquad Let the atom be absorbed on the site $(x,y)$; $\boldsymbol{S}$ get updated. \qquad \emph{\# Absorptions of atoms}\\
			\qquad \qquad \textbf{for} $j$ in $range(0,l)$ \textbf{do:}\\
			\qquad \qquad \qquad Investigate all nearest sites of the adatom on current site: \\
			\qquad \qquad \qquad \textbf{if} Upward diffusion would occur from current site to one targeted site \textbf{then:}\\
			\qquad \qquad \qquad \qquad Do: Energy barrier from current site to the targeted site $\leftarrow$ $+\infty$\\
			\\
			\emph{\#Here, the "$+\infty$" means the diffusion from lower layer to upper layer is forbidden. In practical terms this kind of diffusion }\\
			\emph{\# are not frequent and even rarer under EW effect. Therefore, this approximation will not affect the acurracy of results} \\
			\\
			\qquad \qquad \qquad \textbf{else if} Intralayer diffusion would occur from current site to one targeted site \textbf{then:}\\
			\qquad \qquad \qquad \qquad \textbf{if} Diffusion and scattering force are in the same direction \textbf{then:}\\
			\qquad \qquad \qquad \qquad \qquad Do: Energy barrier from current site to the targeted site $\leftarrow$ $bn1$\\
			\qquad \qquad \qquad \qquad \textbf{else if} the direction of diffusion is perpendicular to that of scattering force \textbf{then:}\\
			\qquad \qquad \qquad \qquad \qquad Do: Energy barrier from current site to the targeted site $\leftarrow$ $bn2$\\
			\qquad \qquad \qquad \qquad \textbf{else if} diffusion and scattering force are in the opposite direction \textbf{then:}\\
			\qquad \qquad \qquad \qquad \qquad Do: Energy barrier from current site to the targeted site $\leftarrow$ $bn3$\\
			\qquad \qquad \qquad \qquad \textbf{end if}\\
			\qquad \qquad \qquad \textbf{else if} Downward diffusion would occur from current site to one targeted site \textbf{then:}\\
			\qquad \qquad \qquad \qquad \textbf{if} Diffusion and scattering force are in the same direction \textbf{then:}\\
			\qquad \qquad \qquad \qquad \qquad Do: Energy barrier from current site to the targeted site $\leftarrow$ $bes1$\\
			\qquad \qquad \qquad \qquad \textbf{else if} the direction of diffusion is perpendicular to that of scattering force \textbf{then:}\\
			\qquad \qquad \qquad \qquad \qquad Do: Energy barrier from current site to the targeted site $\leftarrow$ $bes2$\\
			\qquad \qquad \qquad \qquad \textbf{else if} diffusion and scattering force are in the opposite direction \textbf{then:}\\
			\qquad \qquad \qquad \qquad \qquad Do: Energy barrier from current site to the targeted site $\leftarrow$ $bes3$\\
			\qquad \qquad \qquad \qquad \textbf{end if}\\
			\qquad \qquad \qquad \textbf{end if}\\
			\qquad \qquad \qquad Determine the direction of diffusion base on the barriers to the nextest site, Eq.(11) in the main text  and the\\
			\qquad\qquad\qquad direction of the last diffusion.\\
			\qquad \qquad \qquad Move the adatom to the targeted site; $\boldsymbol{S}$ get updated \qquad \emph{\# Diffusion}\\
			\qquad\qquad\qquad \textbf{if} The adatom collide with other nucleared adatom \textbf{then:}\\
			\qquad\qquad\qquad\qquad \textbf{break} \qquad \qquad\emph{\# nucleation}\\
			\qquad\qquad\qquad \textbf{end if}\\
			
			\qquad \qquad \textbf{end for}\\
			\qquad \textbf{end for}\\
			\qquad\\
			\qquad Output the surface morphology $\boldsymbol{S}$.
		\end{tabular}
	\end{ruledtabular}
\end{table*}

\begin{table*}[h]
	
	\begin{ruledtabular}
		\begin{tabular}{cccccccc}
			\textrm{Electric Field intensity ($V/m$)}&
			\textrm{$bn1$}&
			\textrm{$bn2$}&
			\textrm{$bn3$}&
			\textrm{$bes1$}&
			\textrm{$bes2$}&
			\textrm{$bes3$}\\
			\colrule
			$2.00\times10^{11}$&0.420&0.510&0.600&0.402&0.517&0.630\\
			$1.75\times10^{11}$&0.426&0.496&0.566&0.428&0.523&0.616\\
			$1.50\times10^{11}$&0.432&0.484&0.536&0.460&0.529&0.598\\
			$1.25\times10^{11}$&0.437&0.473&0.509&0.487&0.534&0.582\\
			$1.00\times10^{11}$&0.447&0.470&0.493&0.508&0.539&0.569\\
			Without EW&0.451 &0.451 & 0.451& 0.544 & 0.544 & 0.544\\			
		\end{tabular}
	\end{ruledtabular}
\begin{flushleft}	
	Table 1. Parameters used to generate Fig.4. The unit of each barrier is $eV$ in this table. \\
	Other input parameters we used: $N$ was set as 50000; $T$ was $273K$; $L$ was set as 100; $l$ was set as 30.
\end{flushleft}	
\end{table*}

\section{universality}
In order to clarify the universality of modulating effect of EW, we also calculated the situations of different adatoms, as shown in Fig.1S. We haven’t found any reported research which demonstrates hopping process is dominant in interlayer diffusion, at least for elemental metal thin films. It can be understood from the fact that the hopping process across the steps experiences a configuration of lower coordinate number (CN) than that of exchange process, and thus has a higher energy barrier than exchange process. Even in materials where interlayer diffusion is dominated by hopping process, the energy of adatoms in the growth system would be lowered by the evanescent wave field. Therefore, in general principle, the downward interlayer diffusion must be enhanced by the applied EW field rather than be undermined, no matter in hopping process or in exchange process.

\begin{figure*}
		\includegraphics[width=17cm]{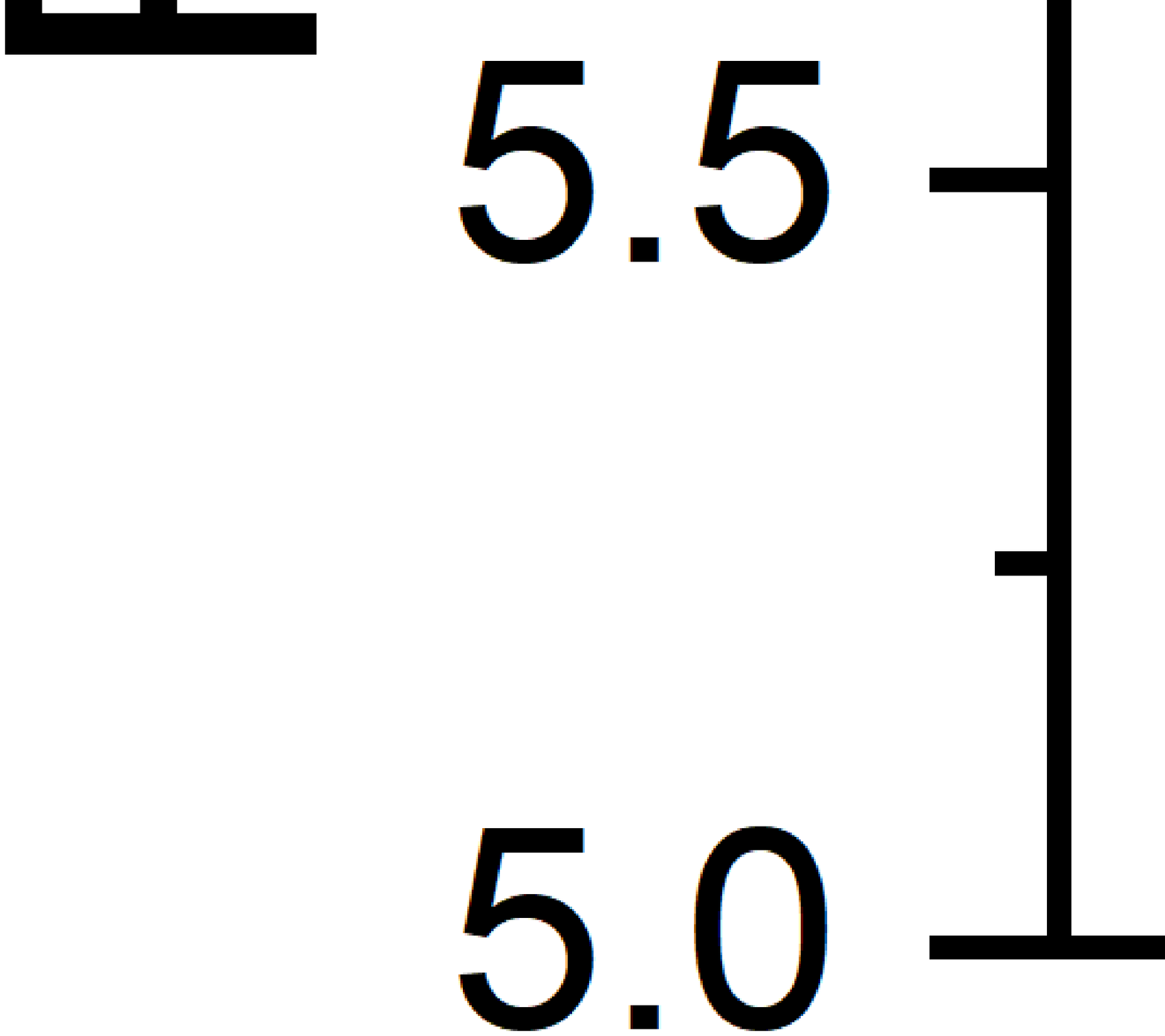}
\begin{flushleft}
	Fig.1S.  Evanescent wave-induced gradient force ($F_{grad}$, marked by solid squares) and scattering force ($F_{scat}$, marked by solid circles) exerted on different metal adatoms in the unit of their respective gravities (G).
\end{flushleft}	
	
\end{figure*}